\documentclass[twocolumn,showpacs,preprintnumbers,amsmath,amssymb]{revtex4}
\usepackage{graphicx}
\usepackage{dcolumn}
\usepackage{bm}
\usepackage{float}
\begin{document}

\title{Measurement-based direct quantum feedback control in an open quantum system}

\author{Yan Yan}
\affiliation{School of Physics, Beijing Institute of Technology, Beijing 100081, People's Republic of China}

\author{Jian Zou}
\email{zoujian@bit.edu.cn}
\affiliation{School of Physics, Beijing Institute of Technology, Beijing 100081, People's Republic of China}

\author{Bao-Ming Xu}
\affiliation{School of Physics, Beijing Institute of Technology,
Beijing 100081, People's Republic of China}

\author{Jun-Gang Li}

\affiliation{School of Physics, Beijing Institute of Technology, Beijing 100081, People's Republic of China}
\author{Bin Shao}

\affiliation{School of Physics, Beijing Institute of Technology, Beijing 100081, People's Republic of China}

\date{Submitted ****}

\begin{abstract}
We consider a general quantum system interacting with a bath and
derive a master equation in the Lindblad form describing the
evolution of the whole quantum system subjected to a
measurement-based direct quantum feedback control (MDFC). As an
example, we consider a qubit coupled with a dephasing environment
under the MDFC. We show that for any given pure target state we can
always find the corresponding MDFC scheme which can effectively
drive any initial state into this target state. And by using
appropriate MDFC scheme with weak measurement we can stabilize a
single qubit initially prepared in one of two nonorthogonal states
against dephasing noise. Furthermore, we can effectively protect a
kind of known mixed states composed of two nonorthogonal states by
using the corresponding MDFC scheme.
\end{abstract}


\maketitle

\section{Introduction}
Quantum information processing tasks often require pure states
as a resource \cite{DiVincenzo2000}, but due to the inevitable interaction between a quantum
system and its surrounding environment, decoherence will happen which
might transform an initial pure state into a statistical mixture.
Because decoherence is the main obstacle for realizing quantum
information tasks, fighting against it has become a major challenge.
Advanced schemes have been proposed to reduce or inhibit decoherence
in quantum system, such as quantum error correction
\cite{Shor1995,Steane1996,Ekert 1996}, decoherence-free subspace
(DFS) \cite{Zanardi1997,Duan1998,Lidar1998,Bacon2000}, dynamical
decoupling (DD) \cite{Agarwal1999,Viola1999,Vitali1999,Xue2013},
engineering reservoir \cite{Poyatos1996}, quantum feedback control
\cite{Wiseman1993,Wiseman1994,Carvalho2007,Carvalho2008,Doherty1999},
etc.
In practice, how to drive an open system from an arbitrary
initial pure or mixed state to a given pure state is very
essential. Usually one begins an experiment with an
unknown state, and then prepares it into a known state, called input
state. Most experiments, including quantum-process-tomography
experiments, require a method to prepare initial states of the
system at the beginning of the experiments
\cite{Alicki1995,Stelmachovic2001,Kuah2007,KuahPhD2007,ModiPhD2008,Modi2011}.
After preparation of
the pure states, keeping high quality of quantum information for a
long time is also crucial for quantum information processing.

As we know steering the dynamics of quantum system by means of
external controls is the central goal in many areas of quantum
physics, especially in quantum information processing. Recent
experimental advances have enabled individual system to be monitored
and manipulated at quantum level in real time
\cite{Hood1998,Julsgard2001,Lu2003,Geremia2004,Ottl2005,Puppe2007},
which makes quantum control more and more practical and realizable.
Among different quantum control schemes quantum feedback control is
widely studied, that is based on feeding back the measurement
results to alter the future dynamics of quantum systems and can be
used to improve the stability and robustness of the system. The
general framework of quantum feedback control was introduced by
Wiseman and Milburn \cite{Wiseman1993,Wiseman1994} leading to
relevant experimental achievements
\cite{Smith2002,Armen2002,Lahaye2004}. In this framework, the
feedback loops in which the measured photocurrent directly
modulating the system Hamiltonian are adopted. The advantage of
Wiseman-Milburn quantum feedback theory is that it is very easy to
consider the limit of Markovian (i.e., instantaneous) feedback and a
master equation of the system can be obtained
\cite{Wiseman1993,Wiseman1994}. Another feedback scheme is Bayesian,
or state estimation quantum feedback \cite{Doherty1999}. It can
result in an improvement over the direct feedback scheme but comes
at the cost of an increasing complexity in experimental
implementation due to the need for a real-time estimation of the
quantum state. Despite being simpler to implement, direct feedback
still exhibits a multitude of possibilities due to the arbitrariness
of choices for control Hamiltonian and measurement scheme. Recently
it has been shown that a direct feedback scheme based on continuous
monitoring of quantum jumps, together with an appropriate choice of
feedback Hamiltonian, can lead to an improvement in amount and
robustness of the steady-state entanglement for two driven and
collectively damped qubits \cite{Carvalho2007}. And it has been also
shown that the jump-based feedback strategy outperforms that based
on homodyne measurement \cite{Carvalho2008}. It is noted that most
recently the measurement-based quantum feedback for an ensemble of
ytterbium($^{171}\mathrm{Yb}$) atoms has been experimentally
implemented \cite{Inoue2013}, and it was shown that the
measurement-based quantum feedback could successfully realize the
unconditional quantum-noise suppression.

Information gain and disturbance in quantum systems are always
antagonistic in quantum theory, and it is argued that a better
feedback control scheme should reach a trade-off between them. A
recently proposed feedback control scheme has introduced weak
measurement in the case of pure dephasing noise \cite{Branczyk2007}.
In Ref.~\cite{Branczyk2007}, a qubit prepared in one of two
nonorthogonal states and subsequently subjected to dephasing noise
was discussed, and it was found that a quantum control scheme based
on a nonprojective measurement with an optimum measurement strength
could realize the optimal recovery from noise. These states are
interesting because due to their nonorthogonality, people think that
it is impossible to design a control procedure that can perfectly
discriminate the input state \cite{Wiseman2009} and subsequently
control the resulting
 input against noise.
And then it was explored from the
experimental point of view to stabilize this nonorthogonal states
against dephasing by using weak measurement-based feedback control
\cite{Gillett2010}.
Later it has been
shown that this scheme of feedback
control based on weak measurement is sensitive to the system state but, for
some suitable states, this scheme works well for four
types of typical noise sources (bit-flip noise, amplitude-damping
noise, phase-damping noise and depolarizing noise) \cite{Xiao2011}.
Moreover, extended techniques for protecting more general nonorthogonal states were presented
in Ref.~\cite{Yang2013}. It is worth noting that in all the above approaches, they
used the Kraus operators to stand for noise and the feedback
control was described by a control map.

In this paper we consider a general quantum system interacting with
a bath and derive a master equation in the Lindblad form describing
the evolution of the open quantum system subjected to a
measurement-based direct quantum feedback control (MDFC). And as an
application of our general master equation, in this paper we
consider a two-level quantum system interacting with a bath of
harmonic oscillators which emulates a dephasing environment.
Different quantum feedback control schemes strongly depend on the
measuring schemes. Being different from Refs.
\cite{Carvalho2007,Carvalho2008} in which the measurement is applied
to the environment (more specifically continuously monitoring the
environment to observe if it absorbs a photon or not), we consider
that the measurement is applied to the open system, that is
described by a positive operator-valued measure (POVM). Based on
superoperator algebra and Nakajima-Zwanzig projectors
\cite{Nakajima1958,Zwanzig1960}, we solve the master equation
numerically. It is found that by using proper MDFC scheme we can
effectively drive any initial pure or mixed state into an arbitrary
given target pure state. Moreover, we find that the optimal feedback
with weak measurement is more effective to protect the system
prepared in one of two nonorthogonal states against decoherence than
the one with projective measurement or do-nothing, that is
consistent with the results of Refs.
\cite{Branczyk2007,Gillett2010}. Finally, we find an effective MDFC
scheme to protect a kind of known mixed states composed of two
nonorthogonal states.

The paper is organized as follows: In Sec. II, we derive a general
master equation in the Lindblad form describing an open quantum
system interacting with a bath under the MDFC. In Sec. III, we
consider a qubit coupled with a dephasing environment
under the MDFC and briefly show how to solve the corresponding master
equation. In Sec. IV, we show that by using appropriate MDFC shemes
we can drive any initial state into an arbitrary given pure state.
In Sec. V, we discuss the effect of the MDFC scheme with weak
measurement to protect the initial state of the qubit prepared in
one of two nonorthogonal states. By using appropriate MDFC scheme we
demonstrate that a kind of known mixed states composed of two
nonorthogonal states can be protected in Sec. VI. Summary and
discussion are given in Sec. VII.

\section{Derivation of a general measurement-based direct feedback control master equation}
In this section, we will derive a master equation describing a
quantum system interacting with a bath under the MDFC. We suppose
that the Hamiltonian of the system is $\hat{H}_{S}$, and the
Hamiltonian of the environment is $\hat{H}_{B}$. The interaction
between the system and its environment is described by
\begin{equation}\label{eq:1}
\hat{H}_{SB}=\sum_{k}\hat{S}_{k}\hat{B}_{k},
\end{equation}
and for each index $k$, $\hat{S}_{k}$ operates only on the system
\textbf{S} and $\hat{B}_{k}$ only on the environment \textbf{B}. The
form of the interaction Eq.~(\ref{eq:1}) is general enough, for both
amplitude and phase damping models \cite{Breuer2002}. We choose to treat the environment and the
interaction between the open system and the environment as parts of
the total Hamiltonian $\hat{H}$,
\begin{equation}\label{eq:2}
\hat{H}=\hat{H}_{S}+\hat{H}_{B}+\hat{H}_{SB}.
\end{equation}

In this paper we choose the POVM measurement, that is described by a
set of measurement operators $\{\hat{M}_{j}\}$ satisfying the
completeness condition $\sum_{j}
\hat{M}_{j}^\dagger\hat{M}_{j}=\hat{I}$. We suppose that the
measurements take place instantaneously and randomly in time, but at
an average rate $R$. After each measurement, we subsequently perform an
instantaneous feedback unitary rotation $\hat{F}_{j}$ based on the
measurement result. Both the measurement and feedback control
act only on the system. We consider a short time interval
$\bigtriangleup t$, during which the probability that a single
measurement as well as the corresponding feedback will occur is
$R\bigtriangleup t$, and we suppose that the time interval of two sequential
measurements or feedbacks is short enough so that the possibility that
two or more measurements and feedbacks occur can be neglected  \cite{Cresser2006}.

The average evolution of the whole system (the open system and the bath) results from adding
both the measurement-based feedback and the normal Schr\"{o}dinger evolution, weighted by their probability of occurrence, to the
lowest order in $\triangle t$, is
\begin{equation}\label{eq:3}
\begin{split}
\hat{\rho}_{SB}(t+\bigtriangleup t)=&(1-R\triangle t)\hat{\rho}_{SB}(t)-\frac{i}{\hbar}[\hat{H},\hat{\rho}_{SB}(t)]\triangle t\\
\\
&+R\triangle
t\sum_{j}(\hat{F}_{j}\hat{M}_{j})\hat{\rho}_{SB}(t)(\hat{F}_{j}\hat{M}_{j})^{\dag}.
\end{split}
\end{equation}
Taking the limit $\bigtriangleup t\rightarrow 0$, we obtain the MDFC
master equation
\begin{equation}\label{eq:4}
\begin{split}
\dot{\rho}_{SB}(t)=&-\frac{i}{\hbar}[\hat{H},\hat{\rho}_{SB}(t)]\\
\\
&+R[\sum_{j}(\hat{F}_{j}\hat{M}_{j})\hat{\rho}_{SB}(t)(\hat{F}_{j}\hat{M}_{j})^{\dag}-\hat{\rho}_{SB}(t)].
\end{split}
\end{equation}
This equation, which is clearly of the Lindblad form, can also be
expressed as:
\begin{equation}\label{eq:5}
\begin{split}
\frac{d}{dt}\hat{\rho}_{SB}(t)=&-\frac{i}{\hbar}[\hat{H},\hat{\rho}_{SB}(t)]+\sum_{j}(\hat{L}^{(S)}_{j}\hat{\rho}_{SB}(t)
\hat{L}^{(S)\dag}_{j}
\\
&-\frac{1}{2}\{\hat{L}^{(S)\dag}_{j}\hat{L}^{(S)}_{j},\hat{\rho}_{SB}(t)\}),
\end{split}
\end{equation}
with $\hat{L}^{(S)}_{j}=\sqrt{R}\hat{F}_{j}\hat{M}_{j}$. The first
term on the right-hand side acting on $\hat{\rho}_{SB}(t)$ is the
Liouvillian superoperator, which accounts for the unitary portion of
the propagation, while the second term, the Lindbladian
superoperator, represents the measurement-based direct feedback
dynamics that acts only on the system. It is noted that our approach
is different from that in Refs.
\cite{Branczyk2007,Gillett2010,Xiao2011,Yang2013}, where they used
the Kraus operator to stand for noise and the feedback scheme
was described by a control map. Here, we treat the environment and
the interaction between the system and the environment as parts of
the total Hamiltonian, and Eq.~(\ref{eq:5}) is the time evolution of
the density matrix of the whole system including the open system and
the bath under the MDFC.

\section{The model and solution}
In this and the following sections, we
consider a two-level quantum system interacting with a bath of
harmonic oscillators
\begin{equation}\label{eq:6}
\begin{aligned}
\hat{H}_{S}&=\hbar\omega_{0}\hat{\sigma}_{z},\\
\hat{H}_{B}&=\hbar\sum_{k}\omega_{k}\hat{b}^{\dag}_{k}\hat{b}_{k},
\end{aligned}
\end{equation}
and phase-damping interaction~\cite{Breuer2002}, that is
\begin{equation}\label{eq:7}
\begin{aligned}
\hat{S}_{k}&=\hbar\hat{\sigma}_{z},\\
\hat{B}_{k}&=g_{k}\hat{b}^{\dag}_{k}+g^{*}_{k}\hat{b}_{k},
\end{aligned}
\end{equation}
where $\omega_{0}$ and $\omega_{k}$ are real constants,
$\hat{b}_{k}$ and $\hat{b}^{\dag}_{k}$ are the annihilation and
creation bath operators, respectively, and $g_{k}$ is a complex coefficient. Let
us define the operator:
\begin{equation}\label{eq:8}
\hat{B}\equiv\sum_{k}\hat{B}_{k}=\sum_{k}(g_{k}\hat{b}^{\dag}_{k}+g^{*}_{k}\hat{b}_{k}),
\end{equation}
and therefore, the interaction can be simplified as:
\begin{equation}\label{eq:9}
\hat{H}_{SB}=\hbar\hat{\sigma}_{z}\hat{B}.
\end{equation}
We suppose that the initial state of the whole system is a product state,
i.e., $\hat{\rho}_{SB}(0)\equiv \hat{\rho}_{S}(0)\otimes
\hat{\rho}_{B}(0)$, and the initial state of the environment is given by
\begin{equation}\label{eq:10}
\begin{aligned}
\hat{\rho}_{B}(0)&=\frac{1}{Z_{B}}\prod_{k}e^{-\hbar\beta\omega_{k}\hat{b}^{\dag}_{k}\hat{b}_{k}},\\
Z_{B}&=\prod_{l}\frac{1}{1-e^{-\hbar\beta\omega_{l}}},
\end{aligned}
\end{equation}
where $\beta=(kT)^{-1}$ represents the inverse temperature of the
bath.

We are interested in the evolution of the system whose
density matrix is given by the partial trace over the
environment,
\begin{equation}\label{eq:11}
\hat{\rho}_{S}=\mathrm{Tr}_{B}\{\hat{\rho}_{SB}\}.
\end{equation}
It should be noted that the MDFC master equation Eq. (5) is valid for both strong and weak interaction between the system and the environment. In the following, we will use the approximate method presented in
Ref.~\cite{Brasil2011} to solve the master equation (\ref{eq:5}) and
to obtain the density matrix of the open system. This
method has been proved to be valid by exact numerical calculations
based on the superoperator-splitting method~\cite{Press2007} for weak coupling
between the system and its environment. So in this paper we only consider the weak interaction between the system and the environment, and for convenience let us first review the method below.

For any density matrix
$\hat{X}$, let us define the superoperators $\hat{\hat{B}}$ and
$\hat{\hat{S}}$ acting on the environment and the system,
respectively,
\begin{equation}\label{eq:12}
\begin{aligned}
\hat{\hat{B}}\hat{X}&\equiv-\frac{i}{\hbar}[\hat{H}_{B},\hat{X}],\\
\hat{\hat{S}}\hat{X}&\equiv-\frac{i}{\hbar}[\hat{H}_{S},\hat{X}]+\sum_{j}\left(\hat{L}^{(S)}_{j}\hat{X}\hat{L}^{(S)\dag}_{j}
-\frac{1}{2}\{\hat{L}^{(S)\dag}_{j}\hat{L}^{(S)}_{j},\hat{X}\}\right),
\end{aligned}
\end{equation}
and the interaction superoperator $\hat{\hat{F}}$, acting on both
Hilbert spaces,
\begin{equation}\label{eq:13}
\hat{\hat{F}}\hat{X}\equiv-\frac{i}{\hbar}[\hat{H}_{SB},\hat{X}].
\end{equation}
In order to solve Eq.~(\ref{eq:5}), we can use the Nakajima-Zwanzig
projector superoperator $\hat{\hat{P}}$, defined as
\begin{equation}\label{eq:14}
\hat{\hat{P}}\hat{X}(t)\equiv\hat{\rho}_{B}(t_{0})\otimes
\mathrm{Tr}_{B}{\hat{X}(t)},
\end{equation}
to obtain the hybrid master equation,
\begin{equation}\label{eq:15}
\frac{d}{dt}[\hat{\hat{P}}\hat{\alpha}(t)]=\int_{0}^{t}dt'[\hat{\hat{P}}\hat{\hat{G}}(t)\hat{\hat{G}}(t')\hat{\hat{P}}\hat{\alpha}(t)],
\end{equation}
where
\begin{equation}\label{eq:16}
\hat{\alpha}(t)=e^{-\hat{\hat{S}}t-\hat{\hat{B}}t}\hat{\rho}_{SB}(t),
\end{equation}
and
\begin{equation}\label{eq:17}
\hat{\hat{G}}(t)=e^{-\hat{\hat{S}}t-\hat{\hat{B}}t}\hat{\hat{F}}e^{\hat{\hat{S}}t+\hat{\hat{B}}t}.
\end{equation}
According to Eq.~(\ref{eq:16}), once $\hat{\alpha}(t)$ is
calculated, $\hat{\rho}_{S}(t)$ can be found by the action of
$e^{\hat{\hat{S}}t}$ on the reduced $\hat{\alpha}(t)$, that is
\begin{equation}\label{eq:18}
\hat{\rho}_{S}(t)=e^{\hat{\hat{S}}t}\mathrm{Tr}_{B}\{\hat{\alpha}(t)\}.
\end{equation}

The quantity $\hat{\hat{P}}\hat{\alpha}(t)$ appears on both sides of
Eq.~(\ref{eq:15}) and we can simplify it into
\begin{equation}\label{eq:19}
\hat{\hat{P}}\hat{\alpha}(t)=e^{-\hat{\hat{S}}t}\hat{\rho}_{S}(t)\hat{\rho}_{B}.
\end{equation}
Then we define the operator:
\begin{equation}\label{eq:20}
\hat{R}(t)\equiv e^{-\hat{\hat{S}}t}\hat{\rho}_{S}(t),
\end{equation}
and obtain
\begin{equation}\label{eq:21}
\hat{\hat{P}}\hat{\alpha}(t)=\hat{R}(t)\hat{\rho}_{B}.
\end{equation}
From Eq.~(\ref{eq:21}), we rewrite $\hat{\hat{P}}\hat{\alpha}(t)$ in
Eq.~(\ref{eq:15}) as $\hat{R}(t)\hat{\rho}_{B}$,
and Eq.~(\ref{eq:15}) can be written as \cite{Brasil2011}
\begin{widetext}
\begin{equation}\label{eq:22}
\begin{aligned}
\frac{d}{dt}[\hat{R}(t)\hat{\rho}_{B}]=&\int_{0}^{t}dt'\int_{0}^{\infty}d\omega
J(\omega)\left\{\coth\left(\frac{\hbar\beta\omega}{2}\right)
\cos[\omega(t-t')]+i\sin[\omega(t-t')]\right\}\otimes\hat{\rho}_{B}\\
&\times\left[e^{-\hat{\hat{S}}t}\hat{\sigma}_{z}(e^{\hat{\hat{S}}(t-t')}\{[e^{\hat{\hat{S}}t'}\hat{R}(t)]\hat{\sigma}_{z}\})-
{e^{-\hat{\hat{S}}t}(e^{\hat{\hat{S}}(t-t')}\{[e^{\hat{\hat{S}}t'}\hat{R}(t)]\hat{\sigma}_{z}\})\hat{\sigma}_{z}}\right]\\
&+\int_{0}^{t}dt'\int_{0}^{\infty}d\omega J(\omega)\left\{\coth\left(\frac{\hbar\beta\omega}{2}\right)\cos[\omega(t-t')]-i\sin[\omega(t-t')]\right\}\otimes\hat{\rho}_{B}\\
&\times\left[e^{-\hat{\hat{S}}t}(e^{\hat{\hat{S}}(t-t')}\{\hat{\sigma}_{z}[e^{\hat{\hat{S}}t'}\hat{R}(t)]\})\hat{\sigma}_{z}-
e^{-\hat{\hat{S}}t}\hat{\sigma}_{z}(e^{\hat{\hat{S}}(t-t')}\{\hat{\sigma}_{z}[e^{\hat{\hat{S}}t'}\hat{R}(t)]\})\right],
\end{aligned}
\end{equation}
\end{widetext}
where $J(\omega)$ is the spectral density of the bath,
\begin{equation}\label{eq:23}
J(\omega)=\sum_{l}|g_{l}|^{2}\delta(\omega-\omega_{l}).
\end{equation}
In this paper, we consider the Ohmic spectral distribution,
\begin{equation}\label{eq:24}
J(\omega)=\alpha\omega e^{-\frac{\omega}{\omega_{c}}},
\end{equation}
where $\alpha\geq0$ is a constant representing the intensity of the
coupling between the system and its environment, and
$\omega_{c}\geq0$ is the cutoff frequency.
Then we apply
$e^{\hat{\hat{S}}t}$ to $\hat{R}(t)$ and obtain the reduced density operator of the system,
\begin{equation}\label{eq:25}
\hat{\rho}_{S}(t)=e^{\hat{\hat{S}}t}\hat{R}(t).
\end{equation}

\section{Preparing an arbitrary given state}
There is a great deal of literature regarding state preparation,
however, many discussions focus on preparing specific
states, e.g., squeezed states, entangled states, etc.
In this section, we will show that we can prepare an arbitrary given pure state from any initial state by using appropriate MDFC
scheme. Without losing generality, we suppose that the arbitrary target state is
$|\psi\rangle_{\texttt{target}}=\cos{\frac{\eta}{2}}|0\rangle+e^{i\zeta}\sin{\frac{\eta}{2}}|1\rangle$
where $\eta\in[0,\pi]$ and $\zeta\in[0,2\pi]$.
We choose the projective measurements $\hat{M}_{+}=|0\rangle\langle0|$ and
$\hat{M}_{-}=|1\rangle\langle1|$. Based on the measurement outcomes ``$\pm$'', we perform
specific correcting rotations. If the outcome is ``$+$'', the corresponding
feedback rotation $\hat{F}_{+}$ is:
\begin{equation}\label{eq:26}
\begin{aligned}
    \hat{F}_{+}=&\exp(-i\zeta \frac{\hat{\sigma}_{z}}{2})\exp(-i\eta\frac{\hat{\sigma}_{y}}{2})\\
          =&\left(\begin{array}{cc}
                                         e^{-i\zeta/2}\cos\frac{\eta}{2} & -e^{-i\zeta/2}\sin\frac{\eta}{2} \\
                                         e^{i\zeta/2}\sin\frac{\eta}{2} & e^{i\zeta/2}\cos\frac{\eta}{2}
                                        \end{array}\right),
\end{aligned}
\end{equation}
and if ``$-$'', the corresponding $\hat{F}_{-}$ is:
\begin{equation}\label{eq:27}
\begin{aligned}
        \hat{F}_{-}=&\exp(-i\zeta \frac{\hat{\sigma}_{z}}{2})\exp(-i(\eta-\pi)\frac{\hat{\sigma}_{y}}{2})\\
          =&\left(\begin{array}{cc}
                                         e^{-i\zeta/2}\sin\frac{\eta}{2} & e^{-i\zeta/2}\cos\frac{\eta}{2} \\
                                         -e^{i\zeta/2}\cos\frac{\eta}{2} & e^{i\zeta/2}\sin\frac{\eta}{2}
                                       \end{array}\right),
\end{aligned}
\end{equation}
where $\hat\sigma_y$ and $\hat\sigma_z$ are the usual Pauli operators. From Eqs.~(\ref{eq:26}, \ref{eq:27}) we can see that $\hat{F}_{+}$ (or $\hat{F}_{-}$)
corresponds to rotating around the $y$ axis of the Bloch sphere with
rotation angle $\eta$ (or $\eta-\pi$) and then around the $z$ axis with
rotation angle $\zeta$.

Next we will show that for any initial state we can
drive it to the target state. An arbitrary
initial state of the qubit can be expressed as,
\begin{equation}\label{eq:28}
  \hat{\rho}_{S}(\textbf{a},0)=\frac{1}{2}(\textbf{I}+\textbf{a}^{T}\cdot{\bm \sigma}),
  \end{equation}
where $\textbf{I}$ is the identity operator,
$\textbf{a}=(|\textbf{a}|\sin{\theta}\cos{\varphi},|\textbf{a}|\sin{\theta}\sin{\varphi},|\textbf{a}|\cos{\theta})^{T}$
is a vector of norm $|\textbf{a}|$ in 3-dimensional real vector
space $\mathbb{R}^{3}$, and ${\bm
\sigma}=(\hat{\sigma}_{x},\hat{\sigma}_{y},\hat{\sigma}_{z})^{T}$ is
a vector with its components being the Pauli operators
$\hat{\sigma}_{j}$ ($j=x,y,z$). The fidelity between the final state
$\hat{\rho}_{S}(\textbf{a},t)$ with initial state parameter
$\textbf{a}$ and the target state $|\psi\rangle_{\texttt{target}}$ is
\begin{equation}\label{eq:29}
  f_{S}(\textbf{a},t)=\sqrt{_{\texttt{target}}\langle\psi| \hat{\rho}_{S}(\textbf{a},t)|\psi\rangle_{\texttt{target}}}.
  \end{equation}
To quantify the universality of our feedback control scheme, we
suppose that the initial state of the qubit system is unknown but
$|\textbf{a}|=\frac{1}{2}$ (see Eq.~(\ref{eq:28})). We define
$F_{S}(t)$ as an average fidelity over all the possible initial states
of Eq.~(\ref{eq:28}) with $|\textbf{a}|=\frac{1}{2}$,
\begin{equation}\label{eq:30}
  F_{S}(t)\equiv\int f_{S}(\textbf{a},t)d\textbf{a},
  \end{equation}
where $d\textbf{a}$ is the (normalized) Haar measure over the
surface of the sphere with radius $|\textbf{a}|$, and
$d\textbf{a}=\frac{1}{4\pi}\sin{\theta}d\theta d\varphi$.

\begin{center}
\includegraphics[scale=0.4]{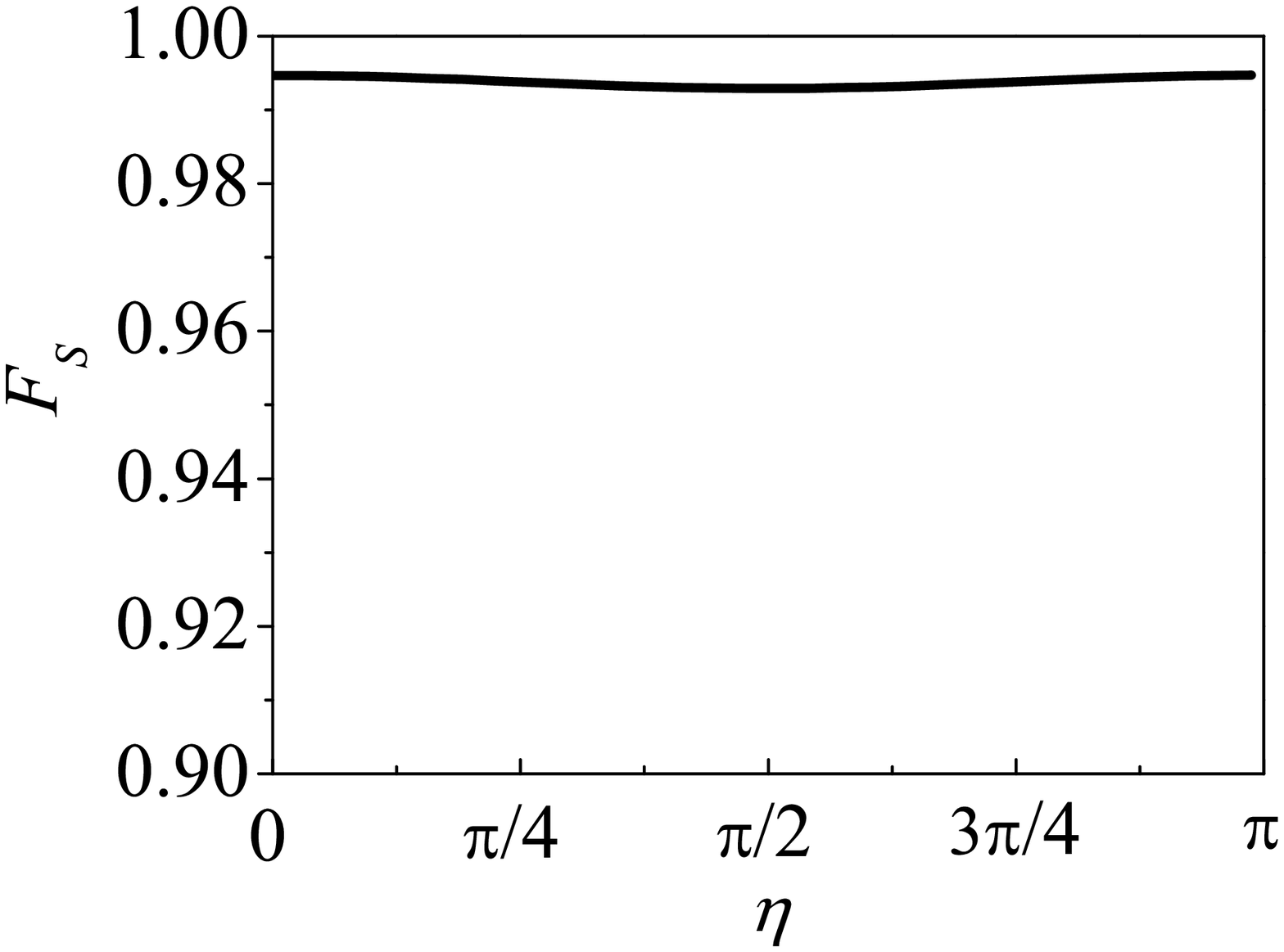}
\parbox{8cm}{\small{\bf Fig. 1}
The average fidelity as a function of the feedback rotation angle $\eta$ at $\omega_{c}t=1$ for $R=4$, $\alpha=0.05$, $\omega_{0}=0$ and $T=0$.}\label{4}\
\end{center}

Following the calculation procedure shown in section III, we can obtain
the average fidelity numerically. From numerical calculations we
find that no matter what initial state the qubit is in, our MDFC scheme can drive an unknown state to the given target state
$|\psi\rangle_{\texttt{target}}=\cos{\frac{\eta}{2}}|0\rangle+e^{i\zeta}\sin{\frac{\eta}{2}}|1\rangle$ with high fidelity.
In Fig.~1 we plot the average fidelity as a function
  of $\eta$ at $\omega_{c}t=1$. When $\eta=\frac{\pi}{2}$ the average fidelity reaches its minimum value, which means that preparing the superposition state $\frac{\sqrt{2}}{2}(|0\rangle+e^{i\zeta}|1\rangle)$ is not as effective as preparing the states
  $|0\rangle$ and $|1\rangle$. It is because that the dephasing noise has no effect on the states $|0\rangle$ and $|1\rangle$, while it will bring the superposition state $\frac{\sqrt{2}}{2}(|0\rangle+e^{i\zeta}|1\rangle)$ into a statistical mixture without the feedback control.
It can be seen from Fig. 1 that our MDFC scheme is very effective
and even for the state
$\frac{\sqrt{2}}{2}(|0\rangle+e^{i\zeta}|1\rangle)$ the fidelity can
still be very high. In fact, our MDFC scheme is just to rotate the
two measurement bases to the same Bloch vector corresponding to the
target state. Through this method, we can drive any initial state to
the target state against decoherence. It works more effectively than
the optimal feedback scheme based on weak measurement to protect a
known state presented in Ref.~\cite{Xiao2011}. Our scheme does not
depend on the premeasurement state but on the postmeasurement state,
while the scheme presented in Ref.~\cite{Xiao2011} depends on both
the postmeasurement and the premeasurement states. Furthermore the
final target state we obtain is a stable one which means that as
time evolves the state of the system finally arrives at the target
state.

It should be noted that the effect of our MDFC preparing scheme
depends on the measurement rate $R$, i.e., the higher the measurement rate,
the purer the final state. Besides, the purity of the final state is
also influenced by the coupling intensity $\alpha$. The purity of
the qubit $\hat{\rho}_{S}(\textbf{a},t)$ at time $t$ is given by
\begin{equation}\label{eq:31}
p_{S}(\mathbf{a},t)=\texttt{Tr}[\hat{\rho}_{S}^{2}(\textbf{a},t)],
\end{equation}
for a completely mixed state, $p_{S}(\mathbf{a},t)=\frac{1}{2}$,
and for a pure state, $p_{S}(\mathbf{a},t)=1$. Without loss of
generality we choose a specific initial state and a specific target
state as an example to illustrate the influence of $R$ and $\alpha$.
In Fig.~2, we plot the purity as a function of the scaled time
$\omega_{c}t$ for different $R$ and $\alpha$, with the target state
$|\psi\rangle_{\texttt{target}}=\frac{\sqrt{2}}{2}(|0\rangle+|1\rangle)$
and the initial state parameters $|\textbf{a}|=\frac{1}{2},
\theta=\frac{\pi}{3},\varphi=0$. From Fig.~2, we can see that the
greater the value of the measurement rate $R$, the purer the state
of the qubit at fixed time $t$, and the faster the speed of the
purification. As we expected that the stronger coupling between the
qubit system and the environment makes the purity lower than that of
the weaker coupling.

\begin{center}
\includegraphics[scale=0.4]{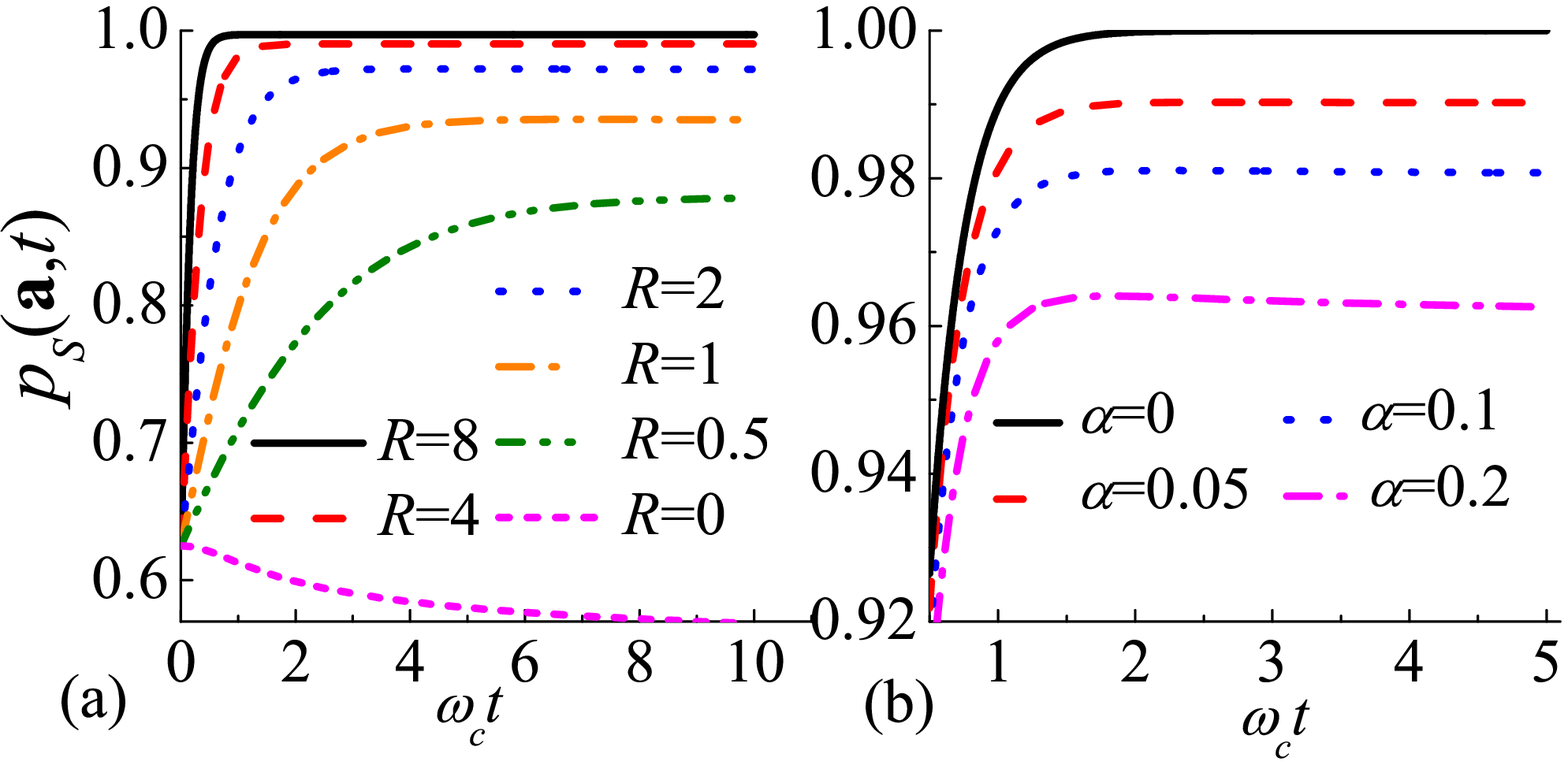}
\parbox{8cm}{\small{\bf Fig. 2}
(Color online) The purity as a function of the scaled time
$\omega_{c}t$ for (a) different measurement rate $R$ ($\alpha=0.05$)
and (b) different coupling strength $\alpha$ ($R=4$). The initial
state parameters $|\textbf{a}|=\frac{1}{2},
\theta=\frac{\pi}{3},\varphi=0$, and the target state
$|\psi\rangle_{\texttt{target}}=\frac{\sqrt{2}}{2}(|0\rangle+|1\rangle)$.
Other parameters are the same as that in Fig.~1.}\label{2}\
\end{center}

\section{PROTECTING TWO NONORTHOGONAL STATES}
Next we consider two nonorthogonal states
\begin{equation}\label{eq:32}
  |\psi_{\pm}\rangle=\cos{\frac{\theta}{2}}|+\rangle\pm\sin{\frac{\theta}{2}}|-\rangle
  \end{equation}
with $|\pm\rangle=\frac{1}{\sqrt{2}}(|0\rangle\pm|1\rangle)$. The
corresponding density matrices are given by
$\hat{\rho}_{\pm}=|\psi_{\pm}\rangle\langle\psi_{\pm}|$, and the
overlapping of the two nonorthogonal states is
$\langle\psi_{+}|\psi_{-}\rangle=\cos{\theta}$. For our model we try
to find a universal MDFC scheme to protect both states from the
dephasing environment, i.e., without knowing which state the qubit
is in, we will attempt to undo the effect of noise.

As we have already known, for our quantum control
scheme, any measurement that acquires information
about a system will inevitably disturb it uncontrollably.
In particular, it suggests that the use of weak measurement could achieve balance between gaining information
and disturbing the original system. Weak measurements have been studied by Aharonov et al. in the context of postselection and the weak value \cite{Aharonov1988,Aharonov1990}. In this section we specifically consider the following measurement operators in our MDFC
scheme,
\begin{equation}\label{eq:33}
\begin{aligned}
    &\hat{M}_{z+}=\cos\frac{\chi}{2}|0\rangle\langle0|+\sin\frac{\chi}{2}|1\rangle\langle1|=\left(\begin{array}{cc}
            \cos\frac{\chi}{2} & 0 \\
            0 & \sin\frac{\chi}{2}
          \end{array}\right),\\
    &\hat{M}_{z-}=\sin\frac{\chi}{2}|0\rangle\langle0|+\cos\frac{\chi}{2}|1\rangle\langle1|=\left(\begin{array}{cc}
            \sin\frac{\chi}{2} & 0 \\
            0 & \cos\frac{\chi}{2}
          \end{array}\right).
\end{aligned}
\end{equation}
The strength of the measurement is adjustable, and it depends on the choice of the
parameter $\chi\in[0,\pi/2]$. $\cos {\chi}$ ranging from 0 (no
measurement) to 1 (projective measurement), and when $0<\cos
{\chi}<1$ it is called weak measurement. Here in order to avoid misunderstanding we emphasize that the weak measurement used in this paper refers to Eq. (33) with $0<\cos {\chi}<1$.
This weak measurement is also called unsharp measurement and has been discussed in Refs.~\cite{Konrad2004,Audretsch2001, Cresser2006,Branczyk2007,Gillett2010,Xiao2011,Yang2013}.
The experimental realization of the weak measurement Eq. (33) was discussed theoretically in Ref.~\cite{Audretsch2001}, and it was shown that this kind of weak measurement can be realized by coupling the system to a meter and performing the usual projective measurements on the meter only.
The experimental implementation was realized recently in a photonic architecture  \cite{Gillett2010}. The required weak measurement on the signal qubit (photon) is realized by entangling it to another meter qubit (photon), and then a full strength projective measurement is performed on the meter qubit, which implements a measurement on the signal qubit with a strength (ranging from do-nothing $\cos \chi =0$ to projective measurement $\cos \chi =1$) determined by the input meter state. A measurement of this kind
can also be realized in nuclear magnetic resonance by means of coupling the spin under consideration to one of its neighbours
\cite{Pegg1989}. Based on the measurement outcomes $z\pm$ of Eq.~(\ref{eq:33}), we then perform correction rotations $\hat{Y}_{\pm
\eta}=\exp(\mp i\eta \hat\sigma_y/2)$ with an angle $\eta$ around
the $y$ axis of the Bloch sphere in a counterclockwise or clockwise
way:
\begin{equation}\label{eq:34}
    \hat{Y}_{\pm\eta}=\exp(\mp\frac{i\eta \hat\sigma_y}{2})=\left(\begin{array}{cc}
                                         \cos\frac{\eta}{2} & \mp\sin\frac{\eta}{2} \\
                                         \pm\sin\frac{\eta}{2} & \cos\frac{\eta}{2}
                                       \end{array}\right).
\end{equation}

Without loss of generality, assuming an
equal probability for sending either state $|\psi_{+}\rangle$ or
$|\psi_{-}\rangle$, we consider the average fidelity $F_{S}(t)$ between the input state and the state at time $t$:
\begin{equation}\label{eq:35}
  F_{S}(t)=\frac{1}{2}\sqrt{\langle\psi_{+}|\hat{\rho}_{+}(t)|\psi_{+}\rangle}+\frac{1}{2}\sqrt{\langle\psi_{-}|\hat{\rho}_{-}(t)|\psi_{-}\rangle}.
  \end{equation}
For any given two nonorthogonal states from numerical calculations
we can find the optimal MDFC schemes for projective measurement and
weak measurement, respectively. As an example we consider that the
two nonorthogonal states are
$|\psi_{\pm}\rangle=\cos{\frac{\pi}{12}}|+\rangle\pm\sin{\frac{\pi}{12}}|-\rangle$.
From numerical calculations we find that for projective measurement,
$\chi=0$, the optimal feedback rotation angle is $\eta=1.3$; for
weak measurement, the optimal parameters are: $\chi=1.0$ and
$\eta=0.5$. In Fig.~3 we plot the fidelity as a function of the
scaled time $\omega_c t$ for three cases: do-nothing; optimal MDFC
scheme with projective measurement; optimal MDFC scheme with weak
measurement. It can be seen from Fig.~3 that the fidelity
improvement by optimal MDFC scheme with weak measurement is better
than that of do-nothing, and surprisingly the effect of the optimal
MDFC scheme with projective measurement is even worse than that of
do-nothing. Our results are consistent with the results in Refs.
\cite{Branczyk2007,Gillett2010}, where the fidelity improvements by
weak measurement also depend on the input states and the largest
improvement is only about $2.5\%$. And it is noted that although we
use Eq.~(\ref{eq:35}) to obtain the optimal MDFC schemes, the
obtained optimal MDFC schemes are universal for both nonorthogonal
states. It is concluded that the optimal MDFC scheme with weak
measurement can protect the initial state a little better than
do-nothing, but most MDFC schemes will accelerate the rate of the
qubit away from its initial state.

\begin{center}
\includegraphics[scale=0.4]{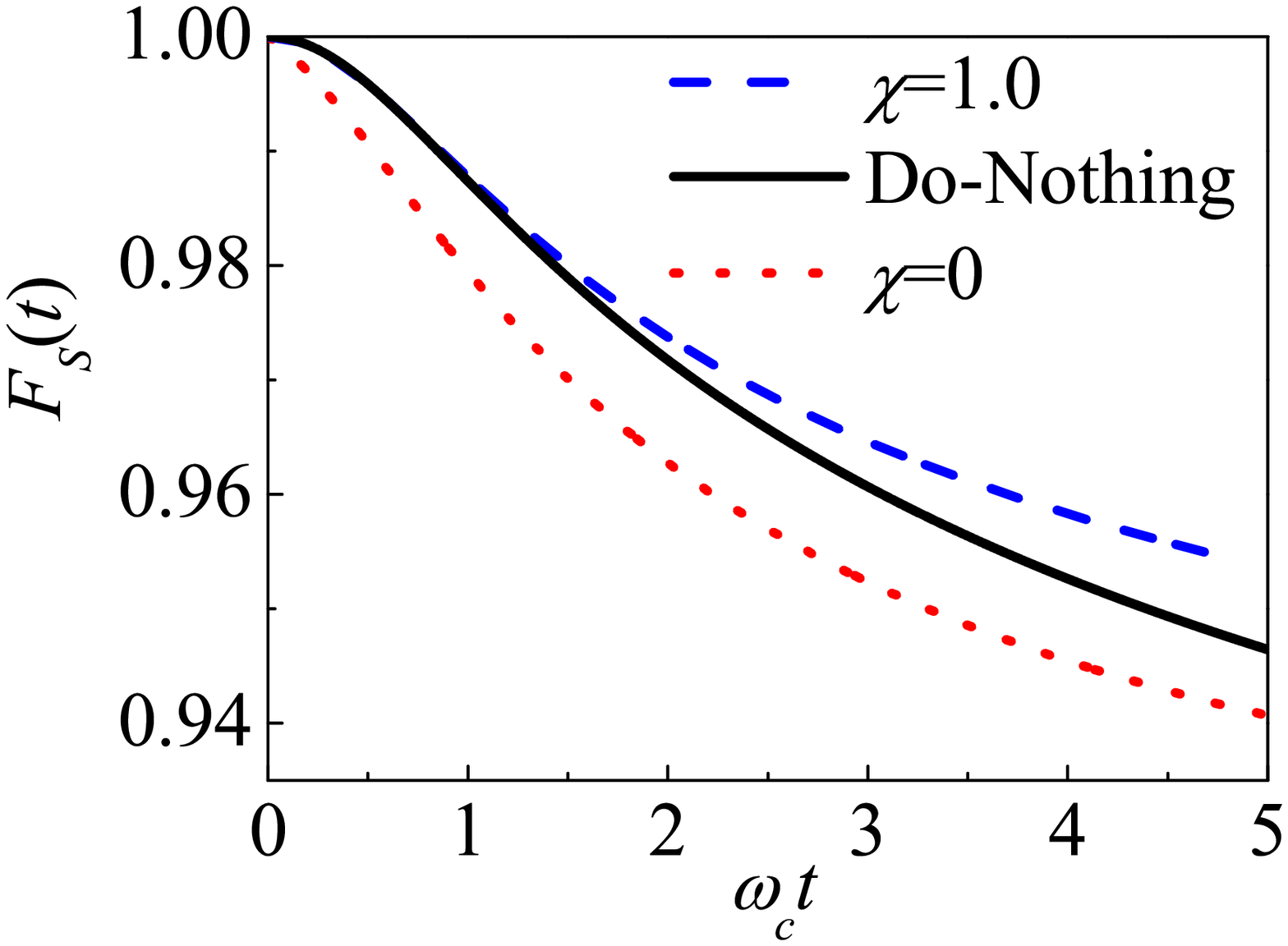}
\parbox{8cm}{\small{\bf Fig. 3}
(Color online) The fidelity as a function of the scaled time $\omega_{c}t$.
The optimal MDFC schemes for projective measurement with $\chi=0$ (red dot line),
weak measurement with $\chi=1.0$ (blue dash line) and without any measurement and
operation (black solid line). The initial parameter is $\theta=\frac{\pi}{6}$. $R=0.5, \alpha=0.05, \omega_{0}=0, T=0$.}\label{3}\
\end{center}

Previous works on quantum state protection using quantum feedback
control have focused on protocols for known states and have not
addressed the issue of protecting unknown quantum states
\cite{Wang2001,Korotkov2001,Ahn2002}. We have just discussed how to
protect two nonorthogonal states against dephasing noise by using
appropriate MDFC scheme. In this case we know what the two states
are and we just do not know which one is sent to us, and we have
found that the scheme with weak measurement is better than that with
projective measurement. Now we investigate whether there exists a
general effective MDFC scheme for unknown quantum states. We assume
that the protected initial state of the system is an arbitrary pure
state as given in Eq.~(\ref{eq:28}) with $|\textbf{a}|=1$. The
fidelity between an initial state
$\hat{\rho}_{S}(\textbf{a},0)\equiv
|\psi(\textbf{a},0)\rangle\langle\psi(\textbf{a},0)|$ with parameter
$\textbf{a}$ and a final state $\hat{\rho}_{S}(\textbf{a},t)$ can be
defined similarly as Eq.~(\ref{eq:29}), that is,
$f_{S}(\textbf{a},t)=\sqrt{\langle\psi(\textbf{a},0)|
\hat{\rho}_{S}(\textbf{a},t)|\psi(\textbf{a},0)\rangle}$, measuring how much two states overlap with each other. In general, the
fidelity varies with the initial states, and if not, the quantum
feedback scheme is said to be universal. We can also define the
average fidelity $F_{S}(t)$ over all the initial states with
different $\theta$ and $\varphi$, $|\mathbf{a}|=1$ for definite MDFC
scheme as in Eq.~(\ref{eq:30}) which quantifies on average how well
the feedback control scheme protect an unknown pure state of the
qubit system against decoherence. In order to find a universal
scheme to protect an unknown quantum state, we calculate the average
fidelity $F_{S}(t)$ for three MDFC schemes, i.e.,
$\hat{L}^{(S)}_{j}=\hat{Y}_{j\eta}\hat{M}_{zj}$ ($j=\pm$),
$\hat{L}^{(S)}_{j}=\hat{Y}_{j\eta}\hat{M}_{xj}$ and
$\hat{L}^{(S)}_{j}=\hat{Z}_{j\eta}\hat{M}_{xj}$, where
\begin{equation}\label{eq:36}
    \hat{M}_{x\pm}=\cos\frac{\chi}{2}|\pm\rangle\langle\pm|+\sin\frac{\chi}{2}|\mp\rangle\langle\mp|,
\end{equation}
with $|\pm\rangle=\frac{\sqrt{2}}{2}(|0\rangle\pm|1\rangle)$,
\begin{equation}\label{eq:37}
    \hat{Z}_{\pm\eta}=\exp(\mp\frac{i\eta \hat{\sigma}_{z}}{2}),
\end{equation}
and $\hat{M}_{z\pm}$ and $\hat{Y}_{\pm\eta}$ are the same as
Eqs.~(\ref{eq:33}, \ref{eq:34}). From numerical calculations we find
that for all the three cases the best schemes are do-nothing, and
all the nontrivial MDFC schemes will accelerate the rate of the
qubit away from its initial state. In one word we could not find a
general effective MDFC scheme to protect an unknown state.

\section{PROTECTING a kind of mixed states composed of two nonorthogonal states}
If we know the qubit is in the two nonorthogonal states $|\psi_{+}\rangle$ and
$|\psi_{-}\rangle$, i.e., Eq.~(\ref{eq:32}), with equal probability, we can describe the system by a density matrix
$\hat{\rho}_{S}(0)=\frac{1}{2}\hat{\rho}_{+}+\frac{1}{2}\hat{\rho}_{-}$. Now we discuss how to protect this mixed state. Due to $\langle0|\hat{\rho}_{S}(0)|0\rangle=\langle1|\hat{\rho}_{S}(0)|1\rangle=\frac{1}{2}$, we choose the projective measurements $\hat{M}_{+}=|0\rangle\langle0|$ and $\hat{M}_{-}=|1\rangle\langle1|$, and the corresponding feedback rotations
are $\hat{Y}_{\pm\eta}=\exp(\mp\frac{i\eta \hat\sigma_y}{2})$.
The fidelity between the initial density matrix and the final density matrix $\hat{\rho}_{S}(t)$ is defined as following:
\begin{equation}\label{eq:38}
  F_{S}(t)=\texttt{tr}\sqrt{\sqrt{\hat{\rho}_{S}(0)}\hat{\rho}_{S}(t)\sqrt{\hat{\rho}_{S}(0)}}.
  \end{equation}
which is a measure of the effect for protecting the mixed state $\hat{\rho}_{S}(0)$.
\begin{center}
\includegraphics[scale=0.4]{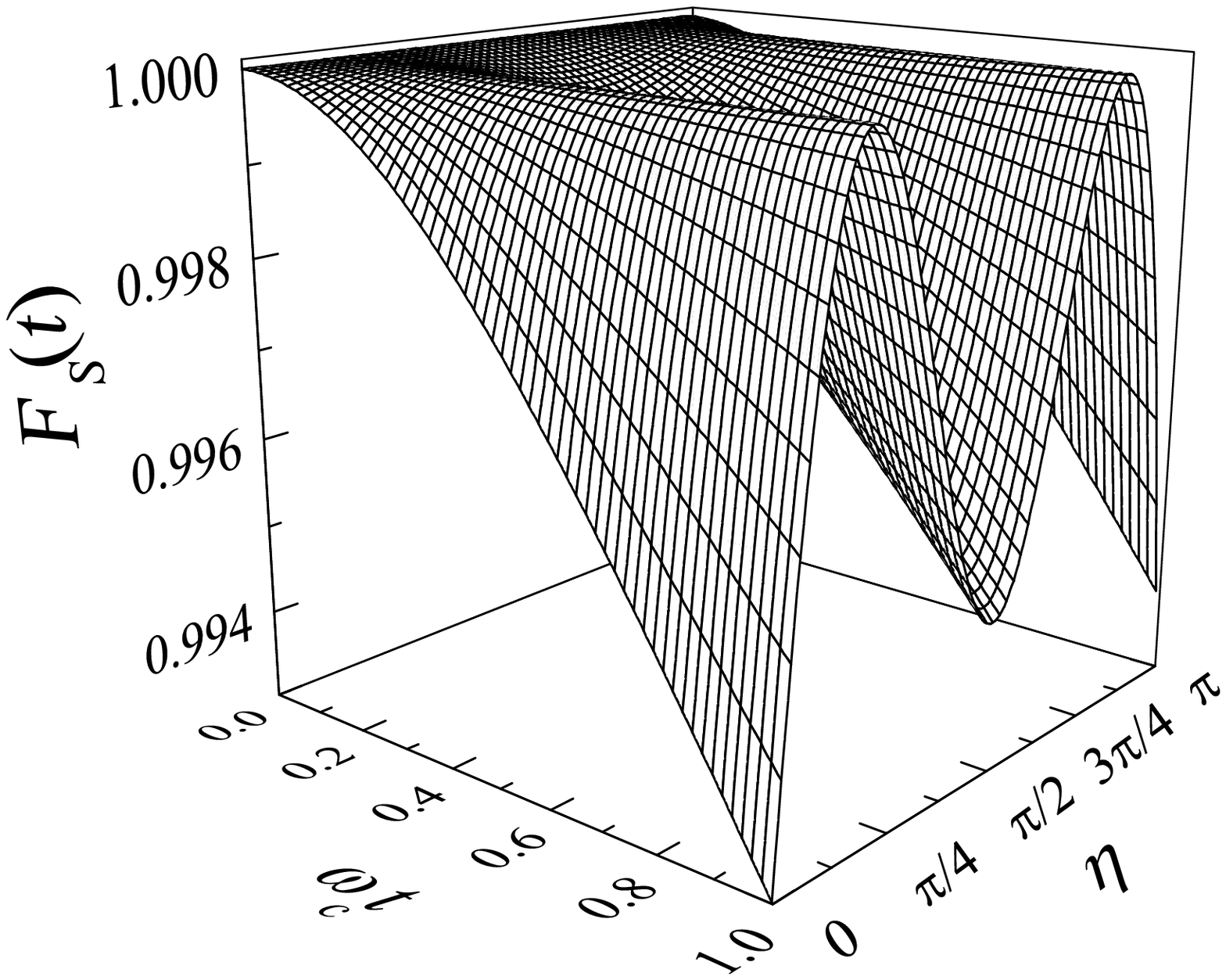}
\parbox{8cm}{\small{\bf Fig. 4}
The fidelity $F_{S}(t)$ as a function of the feedback rotation angle
$\eta$ and the scaled time $\omega_{c}t$. The initial parameter
$\theta=\frac{\pi}{3}$. $R=0.5, \alpha=0.05, \omega_{0}=0,
T=0$.}\label{4}\
\end{center}

From numerical calculations, we plot the fidelity $F_{S}(t)$ as a
function of the feedback rotation angle $\eta$ and the scaled time
$\omega_{c}t$, with the initial parameter $\theta=\frac{\pi}{3}$ as
shown in Fig.~4. It is clear that as time evolves, the density
matrix deviates from its original state without the feedback
($\eta=0$). But for some certain feedback angles
($\eta=\frac{\pi}{6}$ and $\eta=\frac{5\pi}{6}$), the initial state
of the qubit can be protected perfectly. The relationship between
the optimal feedback angle $\eta_{opt}$ and the initial state
parameter $\theta$ can be illustrated in Fig.~5. It is shown that
the optimal feedback angle $\eta_{opt}$ and the initial state
parameter $\theta$ satisfy the relationship
$\eta_{opt}=\frac{\pi}{2}\pm\theta$ which can be seen from Fig.~5.
Our MDFC scheme is like this: we first measure the input state with
the bases $\{|0\rangle$, $|1\rangle\}$, and the state of the qubit
will collapse into $|0\rangle$ or $|1\rangle$ with equal
probability, and then the correction feedback is just rotating the
two states $|0\rangle$ or $|1\rangle$ back to $|\psi_{+}\rangle$ or
$|\psi_{-}\rangle$, respectively.

\begin{center}
\includegraphics[scale=0.4]{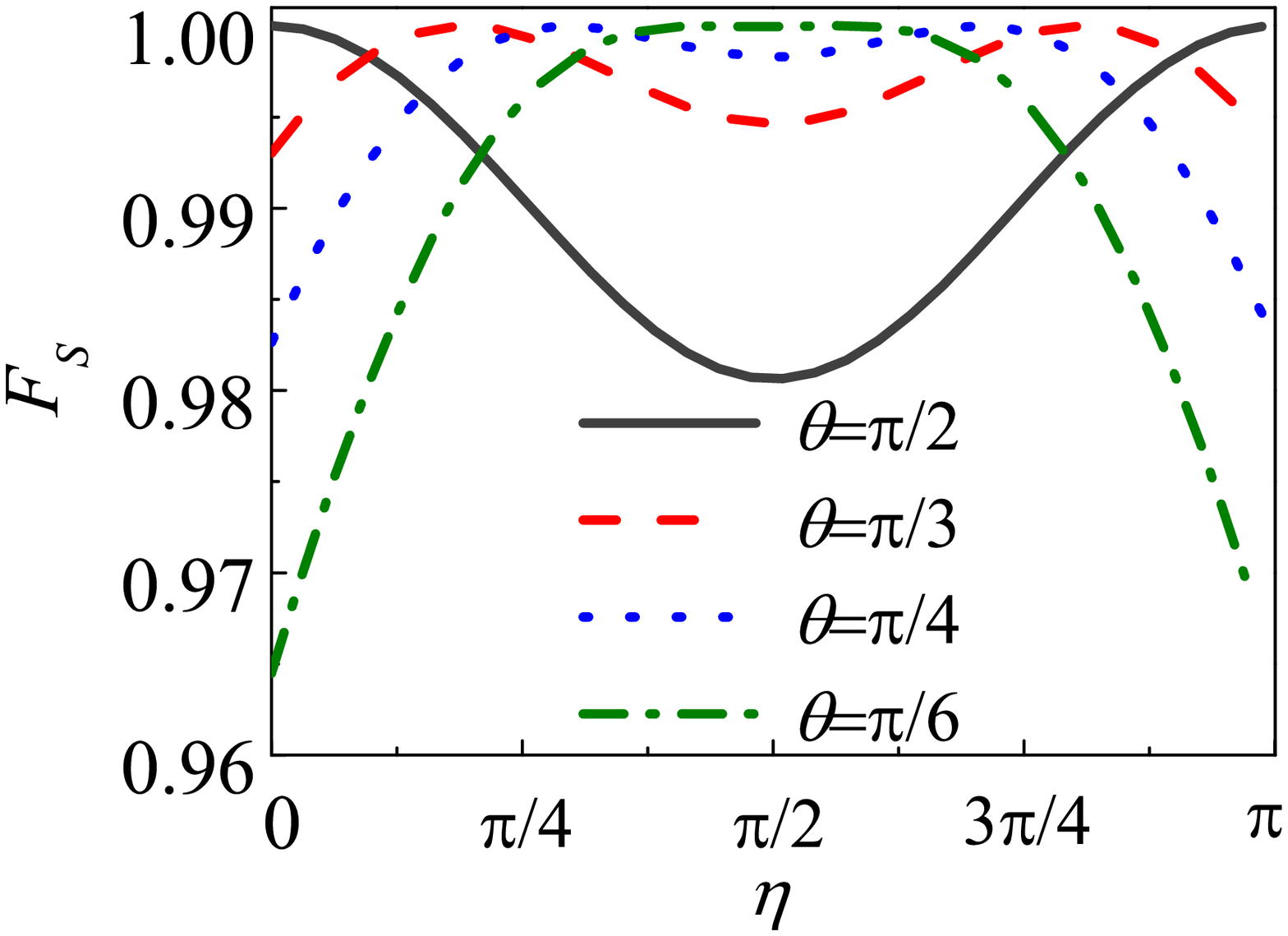}
\parbox{8cm}{\small{\bf Fig. 5}
(Color online) The fidelity $F_{S}$ as a function of the feedback
rotation angle $\eta$ at $\omega_{c}t=1$, for different $\theta$.
$\theta=\frac{\pi}{2}$ (black solid line), $\theta=\frac{\pi}{3}$
(red dash line), $\theta=\frac{\pi}{4}$ (blue dot line),
$\theta=\frac{\pi}{6}$ (olive dash dot line). $R=0.5, \alpha=0.05,
\omega_{0}=0, T=0$.}\label{5}\
\end{center}

Now we consider a general input mixed state composed of two nonorthogonal states $\hat{\rho}_{S}(0)=P_{+}\hat{\rho}_{+}+P_{-}\hat{\rho}_{-}$,
where $P_{\pm}>0$ and $P_{+}+P_{-}=1$. For this state we could find a set of orthogonal measurements
$\hat{M}_{\mu\pm}=|\pm\mu\rangle\langle\pm\mu|$ ($|+\mu\rangle=\cos\frac{\mu}{2}|0\rangle+\sin\frac{\mu}{2}|1\rangle$;
$|-\mu\rangle=\sin\frac{\mu}{2}|0\rangle-\cos\frac{\mu}{2}|1\rangle$) satisfying
\begin{equation}\label{eq:39}
\langle\pm\mu|\hat{\rho}_{S}(0)|\pm\mu\rangle=P_{\pm}.
\end{equation}
After the measurement, the mixed state composed of two nonorthogonal
states is driven to the two bases $|\pm\mu\rangle$ with the
probabilities $P_{\pm}$. Based on the results ``$\pm\mu$'', we
perform the rotations
$U_{|\pm\mu\rangle\rightarrow|\psi_{\pm}\rangle}$ to drive the
density matrix as close as to the input one. In one word by using our
MDFC scheme, we can perfectly stabilize a kind of mixed states
composed of two nonorthogonal states.

\section{SUMMARY AND DISCUSSION}
In this paper, we have derived a general master equation in the
Lindblad form describing the evolution of an open quantum system
subjected to a measurement-based direct quantum feedback control
(MDFC). As an example, we have considered a two-level quantum system
interacting with a dephasing environment under the MDFC. Different
quantum feedback control schemes strongly depend on the measuring
schemes, and in this paper we have considered a general measurement
on the qubit system described by a POVM.
We have found that by using
proper MDFC scheme we can effectively drive any initial pure or
mixed state into an arbitrary given target pure state. It is noted
that our scheme of preparing a quantum state is not to prepare a
particular state at a given time but to prepare an arbitrary given
target pure state from any initial pure or mixed state and
stabilize it against decoherence. For the initial state prepared in
one of two nonorthogonal states we have found that the optimal
feedback with weak measurement is more effective to protect the
system against decoherence than the one with projective measurement
and do-nothing, which is consistent with the results of Refs.
\cite{Branczyk2007,Gillett2010}. But we have not found a universal
MDFC scheme to protect an unknown initial state. Finally, we have
demonstrated that by using optimal MDFC scheme we can stabilize a
kind of known mixed states composed of two nonorthogonal states
against dephasing noise.

When we consider a quantum measurement, we usually concern either
obtaining information of the initial state or the state after the
measurement which is in general unpredictable. Generally a
protecting scheme concerns more about the premeasurement state while
a preparing scheme concerns more about the postmeasurement state.
Our MDFC preparing scheme in Sec.~IV belong to the latter. We have
found that for protecting a known state driving directly to the
target state like our MDFC preparing scheme works much better than
the optimal weak measurement scheme shown in \cite{Xiao2011}, that
concerns both the premeasurement and the postmeasurement states.
Actually for a known state we only need to know the postmeasurement
state, because we have already known precisely the protected state.
For a universal MDFC scheme to protect two known nonorthogonal
states the weak measurement-based MDFC scheme is better. In this
case although we know the two nonorthogonal states completely but we
do not know which one is sent to us. So both the premeasurement
state and the postmeasurement state should be concerned, i.e., we
should choose weak measurements to achieve a balance between gaining
information and disturbing the original system. For a given mixed
state composed of two known nonorthogonal states we can design a
MDFC scheme based on projective measurement to effectively protect
it from the dephasing environment. It can be concluded that if we
know the initial state definitely, no matter mixed or pure, the MDFC
scheme based on projective measurement is much better, and if we
do not completely know the the state protected, the MDFC scheme
based on weak measurement is more effective. In fact, it is very
meaningful for protecting an unknown state, but unfortunately from
our study we have not found a universal effective MDFC scheme to
protect an unknown state. We argue that this is reasonable because
we do not know any information about the state of the qubit before
the measurement, and after measurement we just get parts of the
information about the state, and then we do not know how to do
exactly.

Finally we discuss the feasibility of experimental realization. The feedback control of quantum systems using weak measurements have been realized experimentally in a photonic architecture [37], and the MDFC scheme proposed in this paper can be experimentally realized similarly. In Ref. [37] a signal qubit which is encoded in the polarization of single photon, passes through a dephasing noise channel, and the required variable-strength measurement (including strong and weak measurement) on the signal qubit is realized by entangling it to another meter qubit (photon) using a nondeterministic linear optic controlled-Z (CZ) gate, and then a full strength projective measurement on the meter qubit is performed. This implements a measurement on the signal qubit with a strength determined by the input meter state. And finally the outcome of the projective measurement determines the correction rotation on the signal qubit. For our MDFC scheme we can send the output corrected signal qubit back as the input qubit and begin a new process. If this process is repeated continually, the corresponding feedback control scheme is just our MDFC scheme, i.e., in this way our MDFC scheme can be realized experimentally.

\begin{acknowledgments}
This work was supported by the National Natural Science Foundation
of China (Grants No. 11274043, 11075013, and 11005008).
\end{acknowledgments}

\bibliography{SFDM5BIBR2}

\end{document}